\documentclass{PoS}
\usepackage{amsmath}
\usepackage{amssymb}
\usepackage{graphicx}

\makeatletter

\title{Large-scale computation of the exponentially expanding universe in a simplified Lorentzian type IIB matrix model}

\ShortTitle{The exponentially expanding universe 
in a simplified Lorentzian type IIB matrix model}

\author{\speaker{Yuta Ito}\thanks{KEK-TH-1879}\\
        Department of Particles and Nuclear Physics,\\
        Graduate University for Advanced Studies (SOKENDAI),\\
        Tsukuba, Ibaraki 305-0801, Japan\\
        E-mail: \email{yito@post.kek.jp}}

\author{Jun Nishimura\\
        KEK Theory Center, 
        High Energy Accelerator Research Organization,\\
        Tsukuba, Ibaraki 305-0801, Japan\\
        Department of Particles and Nuclear Physics,\\
        Graduate University for Advanced Studies (SOKENDAI),\\
        Tsukuba, Ibaraki 305-0801, Japan\\
        E-mail: \email{jnishi@post.kek.jp}}

\author{Asato Tsuchiya\\
        Department of Physics, Shizuoka University,\\
        836 Ohya, Suruga-ku, Shizuoka 422-8529, Japan\\
        E-mail: \email{tsuchiya.asato@shizuoka.ac.jp}}

\abstract{
The type IIB matrix model is a conjectured nonperturbative 
formulation of superstring theory. Recent studies on the Lorentzian version 
of the model have shown that only three out of nine spatial directions 
start to expand after some critical time. On the other hand, 
due to the unbounded action of the Lorentzian model,
one has to introduce infrared (IR) cutoffs 
in order to make the partition function finite.
In this work we investigate whether the effects of the IR cutoffs 
disappear in the infinite volume limit.
For that purpose, we study a simplified model with large matrix size
up to $N=256$ by Monte Carlo simulation.
First we confirm the exponentially
expanding behavior of the ``universe''.
Then we generalize the form of the IR cutoffs by one parameter,
and find that the results become universal in some region of the parameter.
It is suggested that the effects of IR cutoffs disappear 
in this region, which is confirmed also from the studies of
Schwinger-Dyson equations.
}

\FullConference{The 33rd International Symposium on Lattice Field Theory\\
14 -18 July 2015\\
Kobe International Conference Center, Kobe, Japan*}

\makeatother

\begin{document}

\section{Introduction}

Superstring theory is a promising candidate of the quantum theory of 
gravity. Since it requires the space-time to have ten dimensions,
it is important to understanding how to 
get a four-dimensional space-time from the theory.
A conventional approach is to 
compactify
six dimensional space using various manifolds. 
However, it is known that such compactifications
give rise to many possible vacua 
described by various low energy effective theories in 4d
with different gauge symmetries. 
Since there is no guiding principle to pick up one particular
vacuum as far as one deals with the theory perturbatively,
one definitely needs to consider the theory non-perturbatively.

The type IIB matrix model \cite{Ishibashi:1996xs} was proposed as
a non-perturbative formulation of superstring theory. The important
feature of this model is that the 10d space-time is described dynamically
as the eigenvalue distribution of ten bosonic matrices. Therefore
one can investigate what kind of space-time emerges in this model.
Until recently, this model has been studied after the Wick rotation
since it is known that the partition function of the Euclidean model
is finite without any cutoffs. However, the Euclidean model is not
suitable for studying the real time dynamics because the time coordinate
is treated as purely imaginary. Moreover, a recent study of the Euclidean
model using the Gaussian expansion method indicates that the emergent
space-time seems to favor three dimensions rather 
than four dimensions \cite{Nishimura:2011xy}.

On the other hand, the Lorentzian version of the type IIB matrix model
has been studied by Monte Carlo simulation
in ref.\cite{Kim:2011cr}. Unlike the Euclidean case,
one has to introduce the infrared (IR) cutoffs in both temporal and
spatial directions in order to make the partition function finite.
Despite such subtleties, one can extract the time evolution
of the space-time from configurations generated by Monte
Carlo simulation. Remarkably, it turned out that the SO(9) rotational
symmetry of the 9d space is spontaneously broken down to SO(3) at
some critical time, after which only three out of nine spatial
directions start to expand.

In order to probe further the time evolution of the emergent 3d space,
one needs to increase the matrix size, which makes the Monte Carlo
simulation more and more time-consuming. For this reason,
two simplified models have been studied.
These models are expected to capture
the qualitative behaviors of the expanding space 
at early times \cite{Ito:2013qga,Ito:2013ywa}
and at late times \cite{Ito:2015mxa},
respectively, in the original model.
The results suggested a scenario for the original model that
the exponential expanding behavior of 3d space continues 
for a while at early times and it changes into
a power-law expansion at late times \cite{Ito:2015mxa}.
Thus, in this model, the inflationary universe
seems to be naturally realized.
Moreover, the suggested power-low expansion at late times
seems to be
consistent with the $t^{1/2}$ behavior, 
which is reminiscent of the behavior of
the Friedman-Robertson-Walker universe in the radiation dominated
era.

Let us emphasize here that
the interesting properties 
such as the SSB of SO(9) and the exponential expansion
are obtained after regularizing the model by introducing the
IR cutoffs.
Therefore it is important to examine whether the effects of the IR
cutoffs disappear in the infinite volume limit. 
This is highly nontrivial
since the space-time emerges dynamically in this model. 
In order to see the trends in the infinite volume limit,
we need to use large matrix size,
and hence we consider the simplified model for early times 
in this study.
We generalize the form of the IR cutoffs by one parameter,
and find that the results become universal in some region of the parameter.
It is therefore suggested that the effects of IR cutoffs disappear 
in this region, which is confirmed also from the studies of
Schwinger-Dyson equations.

The rest of this article is organized as follows. 
In section \ref{sec:def-Lorentzian}
we briefly
review the definition of the Lorentzian type IIB matrix model.
In section \ref{sec:def-simplified}
we define the simplified model, which enables us to achieve
large matrices in Monte Carlo simulation, 
and show, in particular, the exponential expansion of space
in this model.
In section \ref{sec:Dependence-on-the} we discuss
the expanding behavior of space-time 
for different choices of the parameter in the IR cutoffs.
Section \ref{sec:Summary-and-discussions}
is devoted to a summary and discussions.

\section{Lorentzian version of the type IIB matrix model}
\label{sec:def-Lorentzian}

The action of the Lorentzian type IIB matrix model \cite{Ishibashi:1996xs}
is given by
\begin{eqnarray}
S_{{\rm b}} & = & -\frac{1}{4g^{2}}{\rm Tr}\left(\left[A_{\mu},A_{\nu}\right]\left[A^{\mu},A^{\nu}\right]\right) \ ,
\label{eq:sb}\\
S_{{\rm f}} & = & 
-\frac{1}{2g^{2}}{\rm Tr}
\left(\Psi_{\alpha}\left(\mathcal{C}\Gamma^{\mu}\right)_{\alpha\beta}
\left[A_{\mu},\Psi_{\beta}\right]\right) \ ,
\label{eq:sf}
\end{eqnarray}
where the bosonic matrices $A_{\mu}$
$\left(\mu=0,\ldots,9\right)$
and the fermionic matrices $\Psi_{\alpha}$
$\left(\alpha=1,\ldots,16\right)$
are $N\times N$, traceless and Hermitian. We have used
the 10d gamma-matrices after the Weyl projection 
denoted by $\Gamma^{\mu}$ and the charge conjugation matrix $\mathcal{C}$.
The coupling constant $g$ is just a scale parameter
in this model since it can be absorbed by rescaling $A_{\mu}$ and 
$\Psi_\alpha$ appropriately. 
The indices $\mu$ and $\nu$ are raised and lowered
using the Minkowski metric 
$\eta_{\mu\nu}={\rm diag}\left(-1,1,\ldots,1\right)$.
One can obtain the Euclidean version 
by making the Wick rotation $A_{0}=iA_{10}$,
where $A_{10}$ is supposed to be Hermitian.

The partition function for the Lorentzian version is 
proposed as \cite{Kim:2011cr}
\begin{equation}
Z=\int dA\, d\Psi\, e^{i\left(S_{{\rm b}}+S_{{\rm f}}\right)} \ ,
\label{eq:z1}
\end{equation}
where the factor ``$i$'' in front of the action is motivated from
the fact that the string world-sheet metric should also have a Lorentzian
signature. Since the Lorentzian version deals with a real time instead of
an imaginary time, it is suitable for investigating the real-time
dynamics. Note, however, that the bosonic action \eqref{eq:sb}
can be written as 
\[
S_{{\rm b}}\simeq-2{\rm Tr}
\left(F_{0i}\right)^{2}+{\rm Tr}\left(F_{ij}\right)^{2} \ ,
\]
where we have introduced 
the Hermitian matrices $F_{\mu\nu}=i\left[A_{\mu},A_{\nu}\right]$.
Therefore $S_{{\rm b}}$ is no longer positive semi-definite, and
it is not bounded from below because the two terms in the action have
opposite signs.

In order to make the partition function \eqref{eq:z1} finite, one
needs to regularize it by introducing IR cutoffs in both
temporal and spatial directions, for instance, as
\begin{eqnarray}
\frac{1}{N}{\rm Tr}\left[\left(A_{0}^{2}\right)^{p}\right] 
& \leq & \kappa^{p} \ ,
\label{eq:cutoff_temp}\\
\frac{1}{N}{\rm Tr}
\left[\left(A_{i}^{2}\right)^{p}\right] & \leq & 1 \ .
\label{eq:cutoff_spc}
\end{eqnarray}
While we have used $p=1$ for simplicity
in the previous work \cite{Kim:2011cr,Ito:2013qga,Ito:2013ywa},
it is important to examine whether the obtained results
depend on the choice of $p$ in the infinite volume limit.

The partition function \eqref{eq:z1} can be rewritten into 
a form which allows direct Monte Carlo studies in the following way.
(See appendix A of ref.~\cite{Ito:2013ywa} for details.)
First, by integrating out the fermionic matrices in the
partition function \eqref{eq:z1}, we obtain the Pfaffian
\begin{equation}
\int d\Psi\, e^{iS_{{\rm f}}}={\rm Pf}\,\mathcal{M}\left(A\right),
\end{equation}
which is real unlike in the Euclidean case. Since $S_{{\rm b}}$ 
becomes $r^{4}S_{{\rm b}}$ under rescaling $A_{\mu}\mapsto rA_{\mu}$,
we can think of integrating out the scale factor of  $A_{\mu}$ first. 
Then the phase factor $e^{iS_{{\rm b}}}$
can be converted into a delta function $\delta\left(S_{{\rm b}}\right)$
for sufficiently large matrix size, and we arrive at
\begin{equation}
Z=\int dA\,{\rm Pf}\mathcal{M}\left(A\right)\delta\left(S_{{\rm b}}\right)
\delta\left(\frac{1}{N}{\rm Tr}\left[\left(A_{i}^{2}\right)^{p}\right]-1\right)
\theta\left(\kappa^{p}-\frac{1}{N}{\rm Tr}\left[\left(A_{0}^{2}\right)^{p}\right]\right) \ .\label{eq:z2}
\end{equation}
Here the delta function
and Heaviside step function $\theta\left(x\right)$
represent the IR cutoffs (\ref{eq:cutoff_temp}) and (\ref{eq:cutoff_spc}).

In order to extract the time evolution from configurations 
generated by simulating \eqref{eq:z2}, 
we diagonalize the temporal matrix $A_{0}$ as
\begin{equation}
A_{0}={\rm diag}\left(\alpha_{1},\ldots,\alpha_{N}\right) \ ,
\quad{\rm where\;}\alpha_{1}<\cdots<\alpha_{N} \ ,
\end{equation}
using the SU($N$) symmetry of the model. It turns out that the spatial
matrices $A_{i}$ have a band-diagonal structure in this basis. In
other words, there exists some integer $n$ such that the magnitude
of elements of the spatial matrices $\left(A_{i}\right)_{ab}$ for
$\left|a-b\right|>n$ are much smaller than those for $\left|a-b\right|<n$.
Based on this observation, we may naturally consider $n\times n$
matrices
\begin{equation}
\left(\bar{A}_{i}\right)_{IJ}\left(t\right)=
\left(A_{i}\right)_{\nu+I,\nu+J} \ ,\label{eq:a_block}
\end{equation}
where $I,J=1,\ldots,n$ and $\nu=0,1,\ldots,N-n$, and $t$ is the
time defined by
\begin{equation}
t=\frac{1}{n}\sum_{I=1}^{n}\alpha_{\nu+I}\label{eq:time}
\end{equation}
corresponding to the $n\times n$ matrices $\bar{A}_{i}$. 
We regard $\bar{A}_{i}\left(t\right)$ as the state of the universe
at time $t$. Using \eqref{eq:a_block}, for example, we can define
the extent of the space at time $t$ as
\begin{equation}
R^{2}\left(t\right)=\left\langle \frac{1}{n}{\rm tr}\sum_{i}\left(\bar{A}_{i}\left(t\right)\right)^{2}\right\rangle  \ ,\label{eq:rt}
\end{equation}
where the symbol tr represents a trace over the $n\times n$ block
matrix. We can also define the ``moment of inertia tensor'' as
\begin{equation}
T_{ij}\left(t\right)=\frac{1}{n}{\rm tr}\left(\bar{A}_{i}\left(t\right)\bar{A}_{j}\left(t\right)\right) \ ,\label{eq:inertia_tenor}
\end{equation}
whose nine eigenvalues $\lambda_{i}\left(t\right)$
represent the spatial extent in each of the nine directions at time
$t$.

\section{The simplified model}
\label{sec:def-simplified}

The most time-consuming part of the Monte Carlo simulation
comes from the Pfaffian.
Let us note here that 
the fermionic action \eqref{eq:sf} can be decomposed into two terms as
\begin{equation}
S_{{\rm f}}\propto{\rm Tr}\left(\bar{\Psi}_{\alpha}\left(\Gamma^{0}\right)_{\alpha\beta}\left[A_{0},\Psi_{\beta}\right]\right)
+{\rm Tr}\left(\bar{\Psi}_{\alpha}\left(\Gamma^{i}\right)_{\alpha\beta}\left[A_{i},\Psi_{\beta}\right]\right) \ .
\label{eq:sf_decomposed}
\end{equation}
At early times, 
the second term is less important than the first one
since the expansion of the universe has not proceeded much.
Therefore we may omit the second term as a simplification. 
Then the Pfaffian simplifies to
\begin{equation}
{\rm Pf}\mathcal{M}\left(A\right)\simeq\Delta^{16}\left(\alpha\right)\ ,
\end{equation}
where 
$\Delta\left(\alpha\right)\equiv\prod_{i>j}\left(\alpha_{i}-\alpha_{j}\right)$
is the van der Monde determinant. 
The partition function of the simplified model is therefore given as
\begin{equation}
Z=\int dA_{i}\prod_{m}d\alpha_{m}\,\Delta^{18}\left(\alpha\right)
\delta\left(S_{{\rm b}}\right)\delta\left(\frac{1}{N}{\rm Tr}\left[\left(A_{i}^{2}\right)^{p}\right]-1\right)
\theta\left(\kappa^{p}-
\frac{1}{N}{\rm Tr}\left[\left(A_{0}^{2}\right)^{p}\right]\right) \ ,
\label{eq:simplified}
\end{equation}
where the extra factor $\Delta^{2}\left(\alpha\right)$ comes from
the Fadeev-Popov procedure for the gauge fixing.

\begin{figure}[t]
\centering{}
\includegraphics{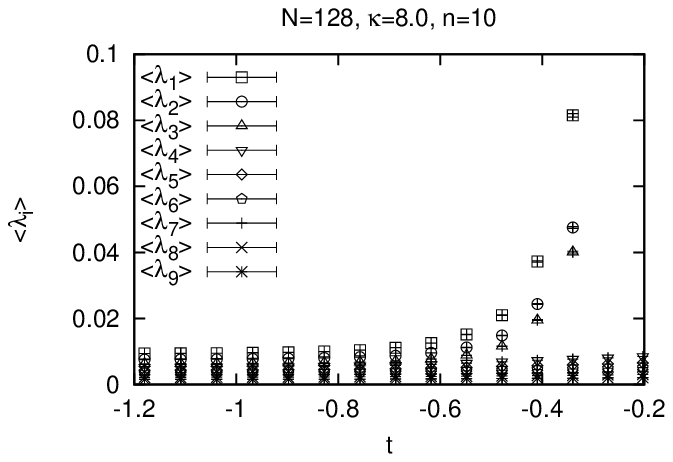}
\includegraphics{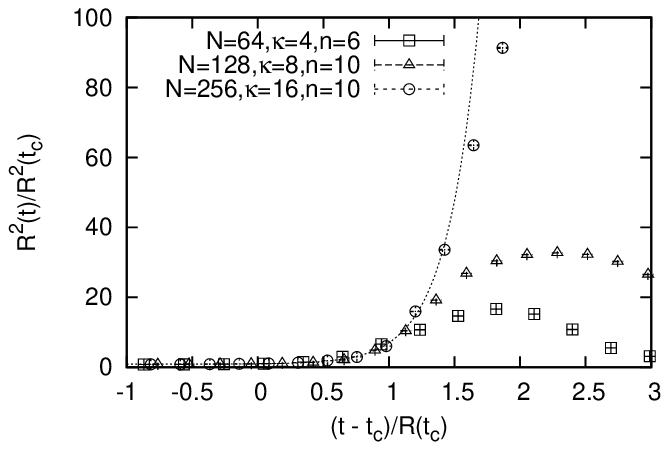}
\caption{(Left) The nine eigenvalues of $T_{ij}\left(t\right)$ are plotted
against time $t$ for the simplified model with $N=128$, $\kappa=8.0$,
$n=10$ using the IR cutoffs (\protect\ref{eq:cutoff_temp}) 
and (\protect\ref{eq:cutoff_spc})
with $p=1$. (Right) The extent of the space $R^{2}\left(t\right)$
normalized by $R^{2}\left(t_{{\rm c}}\right)$ is plotted against
$x=\left(t-t_{{\rm c}}\right)/R\left(t_{{\rm c}}\right)$ for
the simplified model with $N=64$ ,$128$
and $256$ 
using the IR cutoffs (\protect\ref{eq:cutoff_temp}) 
and (\protect\ref{eq:cutoff_spc})
with $p=1$. The dashed line is a fit to 
$R^{2}\left(t\right)/R^{2}\left(t_{{\rm c}}\right)=
a+\left(1-a\right)\exp\left(bx\right)$
with $N=256$ for $0\leq x\leq1.3$, which gives $a=0.89(3)$ and
$b=4.0(3)$.\label{fig:tij_rt}}
\end{figure}

This simplification allows us to study large matrix size.
In Fig.~\ref{fig:tij_rt} (Left),
we plot the expectation values 
$\left\langle \lambda_{i}\left(t\right)\right\rangle $
of the nine eigenvalues of $T_{ij}\left(t\right)$ for the simplified
model with $N=128$ 
using the IR cutoffs (\ref{eq:cutoff_temp}) and (\ref{eq:cutoff_spc})
with $p=1$. We observe a gap
between $\left\langle \lambda_{3}\left(t\right)\right\rangle $ and
$\left\langle \lambda_{4}\left(t\right)\right\rangle $, which
signals the spontaneous symmetry breaking of SO(9) to SO(3). 
We define the critical time $t_{{\rm c}}$ as the time at which
the gap between $\left\langle \lambda_{3}\left(t\right)\right\rangle $ and
$\left\langle \lambda_{4}\left(t\right)\right\rangle $ starts to develop. 
For example, we obtain $t_{{\rm c}}=-0.63108(7)$ for $N=128$
from Fig.~\ref{fig:tij_rt} (Left). Applying the same procedure to
other $N$, we find reasonable large-$N$ scaling behaviors for
the extent of space $R\left(t\right)$.
Considering the ambiguities in defining the critical time 
at finite $N$, we modify the value of $t_{{\rm c}}$ slightly
from the one determined by the above procedure
in such a way that the scaling behavior is optimized.
In Fig.\ref{fig:tij_rt} (Right), we plot
the extent of the space $R\left(t\right)$ obtained in this way.
We find that
the behavior of $R^{2}\left(t\right)$ at $t>t_{{\rm c}}$ can be
fitted to an exponential function.\footnote{The exponential
expansion has been observed
also in the 6d version of the simplified model with
$N \le 64$ \cite{Ito:2013ywa}.
}
It is suggested that the exponential expansion
occurs also in the original Lorentzian type IIB matrix model at early times.
The results obtained 
up to $N=24$ for the original model \cite{Ito:2013qga}
are consistent with this speculation.

\section{Dependence on the IR cutoffs\label{sec:Dependence-on-the}}

In the previous section, 
we have shown the results obtained for the simplified
model using $p=1$ in
the IR cutoffs (\ref{eq:cutoff_temp}) and (\ref{eq:cutoff_spc}).
In this section we discuss how the results
depend on the parameter $p$ in the IR cutoffs.

\begin{figure}[t]
\centering{}
\includegraphics{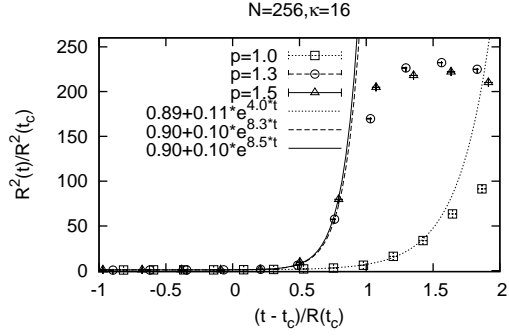}
\caption{The extent of the space 
$R^{2}\left(t\right)$ normalized by $R^{2}\left(t_{{\rm c}}\right)$
is plotted against 
$x=\left(t-t_{{\rm c}}\right)/R\left(t_{{\rm c}}\right)$
with $N=256$ and $\kappa=16$ for $p=1.0$, $1.3$ and $1.5$. 
We have used the block size $n=10$ 
for $p=1.0$ and $n=6$ for $p=1.3$ and $1.5$. 
The 
lines are fits to 
$R^{2}\left(t\right)/R^{2}\left(t_{{\rm c}}\right)=
a+\left(1-a\right)\exp\left(bx\right)$.
\label{fig:rt_various_p}}
\end{figure}

In Fig.~\ref{fig:rt_various_p} we plot $R^{2}\left(t\right)$
against $t$ for various values of $p$.
We find that the results for $p=1.3$ and $1.5$ seem to be 
almost the same except for the region around the peak. 
On the other hand, the result for $p=1.0$
deviates from the others in the whole $t>t_{{\rm c}}$ region. 
These observations suggest that the effects of the IR cutoffs 
disappear for sufficiently large $p$ 
but not for $p=1.0$.

Let us note that, as the power $p$ in
(\ref{eq:cutoff_temp}) and (\ref{eq:cutoff_spc})
is increased,
the IR cutoffs tend to affect only
large eigenvalues of $(A_0)^2$ and $(A_i)^2$, respectively.
This naturally explains why we observe a universal behavior
at sufficiently large $p$.
From this point of view, we may conclude that
the results for $p = 1.3$ and $1.5$
are affected by the IR cutoffs only around the peak,
whereas the results for $p=1.0$ are affected by the IR cutoffs
in the whole time region.

We can fit the results for $p = 1.3$ and $1.5$ by
an exponential function as shown in Fig.~\ref{fig:rt_various_p}.
Thus we find that the result for $p=1.0$ is qualitatively
the same as the ones for $p = 1.3$ and $1.5$.

\section{Summary and discussions\label{sec:Summary-and-discussions}}

In this work, we have studied the issues concerning
the IR cutoffs introduced to regularize 
the Lorentzian type IIB matrix model.
While the model exhibits interesting properties
such as the spontaneous
symmetry breaking of SO(9),
it is important to examine whether the effects of the IR cutoffs 
disappear in the infinite volume limit.

In order to address this issue, we studied
the simplified model for early times,
which is obtained from the original model
by omitting the term proportional
to $A_{i}$ in the fermionic action.
It turned
out that this simplified model also has the properties such as the
SSB of SO(9) down to SO(3).
We have also confirmed the exponential expansion
of space using large matrix size up to $N=256$.

In order to investigate the effects of the IR cutoffs
in such results,
we have introduced the parameter $p$ in \eqref{eq:cutoff_temp}
and \eqref{eq:cutoff_spc}, where $p=1$ corresponds
to the IR cutoffs used so far. 
The qualitative behavior
of $R\left(t\right)$ is the same for all the values of $p$ we studied.
Quantitatively, we find that the results for $p=1.3$ and $1.5$ seem to be
almost the same except for the region near the peak,
whereas the result for $p=1$ is different from the others. 
This can be understood from the fact that
the IR cutoffs tend to affect only the large eigenvalues of
$(A_0)^2$ and $(A_i)^2$ as $p$ is increased.
Thus our results suggest that
the IR cutoff effects disappear in the scaling region
and the results become universal
for sufficiently large $p$ but not for $p=1.0$. 

As a more direct approach to probe the effects of the IR cutoffs,
we have also studied Schwinger-Dyson equations.
According to this analysis, we indeed find that the term arising from
the IR cutoffs becomes smaller as $N$ is increased when 
$p > 1$.
This also suggests that the effects of the IR cutoffs 
disappear in the infinite volume limit for sufficiently large $p$.

Considering that the argument on the disappearance of 
the cutoff effects does not depend on the details of the model, 
we may naturally expect that the cutoff effects
also disappear with sufficiently large $p$ for the original model.
However, the critical value of $p$ beyond which the cutoff effects
disappear in the infinite volume limit may depend on the model. 
It is therefore important to determine the critical $p$
by simulating the original model.

\section*{Acknowledgments}

This research used computational resources of the K computer of the
HPCI system provided by the AICS through the HPCI System Research
Project (Project ID : hp130063, hp150082). The supercomputer FX10 at University
of Tokyo has been used in developing our code for parallel computing.
The work of Y.I. is supported by Grant-in Aid for JSPS fellows.
The work of J.~N.\ and A.~T.\ was supported in part by Grant-in-Aid 
for Scientific Research (No.\ 23244057 for J.~N.\
and No.\ 24540264, 23244057, 15K05046 for A.~T.)
from Japan Society for the Promotion of Science.


\begin{thebibliography}{1}
\bibitem{Ishibashi:1996xs} N.~Ishibashi, 
H.~Kawai, Y.~Kitazawa, and A.~Tsuchiya,
\emph{A large-N reduced model as superstring},
Nucl.\ Phys.\ B, {bf 498}, 467 (1997)
[{\tt arXiv:hep-th/9612115}].

\bibitem{Nishimura:2011xy}
  J.~Nishimura, T.~Okubo and F.~Sugino,
\emph{Systematic study of the 
SO(10) symmetry breaking vacua in the matrix model for type IIB superstrings},
  JHEP {\bf 1110} (2011) 135
  [{\tt arXiv:1108.1293 [hep-th]}].


\bibitem{Kim:2011cr} S.~W.~Kim, J.~Nishimura and A.~Tsuchiya,
\emph{Expanding (3+1)-dimensional universe 
from a Lorentzian matrix model for superstring theory in (9+1)-dimensions},
Phys.\ Rev.\ Lett.\ \textbf{108} (2012) 011601 
[{\tt arXiv:1108.1540 [hep-th]}].


\bibitem{Ito:2013qga} Y.~Ito, S.~W.~Kim, J.~Nishimura and A.~Tsuchiya,
\emph{Monte Carlo studies on the expanding behavior 
of the early universe in the Lorentzian type IIB matrix model},
PoS LATTICE \textbf{2013} (2014) 341 
[{\tt arXiv:1311.5579 [hep-lat]}].


\bibitem{Ito:2013ywa} Y.~Ito, S.~W.~Kim, Y.~Koizuka, J.~Nishimura and A.~Tsuchiya,
\emph{A renormalization group method 
for studying the early universe in the Lorentzian IIB matrix model},
PTEP \textbf{2014} (2014) 8, 083B01
[{\tt arXiv:1312.5415 [hep-th]}].


\bibitem{Ito:2015mxa}
  Y.~Ito, J.~Nishimura and A.~Tsuchiya,
\emph{Power-law expansion of the Universe 
from the bosonic Lorentzian type IIB matrix model},
  JHEP {\bf 1511} (2015) 070
[{\tt arXiv:1506.04795 [hep-th]}].

\end{thebibliography}
\end{document}